# Structure of Water at Hydrophilic and Hydrophobic Interfaces: Raman Spectroscopy of Water Confined in Periodic Mesoporous (Organo)Silicas

Benjamin Malfait[1], Alain Moréac[1], Aïcha Jani[1], Ronan Lefort[1], Patrick Huber[2,3,4,a], Michael Fröba[5,b], and Denis Morineau[1,c]

[1]Institute of Physics of Rennes, CNRS-University of Rennes 1, UMR 6251, F-35042 Rennes, France

[2]Institute for Materials and X-Ray Physics, Hamburg University of Technology, 21073 Hamburg, Germany

[3]Centre for X-Ray and Nano Science CXNS, Deutsches Elektronen-Synchrotron DESY, 22603 Hamburg, Germany

[4]Centre for Hybrid Nanostructures CHyN, Hamburg University, 22607 Hamburg, Germany

[5]Institute of Inorganic and Applied Chemistry, University of Hamburg, 20146 Hamburg, Germany

[a]patrick.huber@tuhh.de

[b]froeba@chemie.uni-hamburg.de

[c]denis.morineau@univ-rennes1.fr

**ABSTRACT:**

The temperature dependence of the structure of water confined in hydrophilic mesostructured porous silica (MCM-41) and hydrophobic benzene-bridged periodic mesoporous organosilicas (PMO) is studied by Raman vibrational spectroscopy. For capillary filled pores (75% relative humidity, RH), the OH-stretching region is dominated by the contribution from liquid water situated in the core part of the pore. It adopts a bulk-like structure that is modestly disrupted by confinement and surface hydrophobicity. For partially filled pores (33% RH), the structure of the non-freezable adsorbed film radically differs from that found in capillary filled pores. A first remarkable feature is the absence of the Raman spectral fingerprint of low density amorphous ice, even at low temperature (-120°C). Secondly, additional bands reveal water hydroxyls groups pointing towards the different water/solid and water/vapor interfaces. For MCM-41, they correspond to water molecules acting as weak H-bond donors with silica, and dangling hydroxyl groups oriented towards the empty center of the pore. For benzene-bridged PMO, we found an additional type of dangling hydroxyl groups, which we attribute to water at hydrophobic solid interface.

Topics: Nanoconfinement – Physisorption – Molecular Dynamics – Water – Liquid solid interfaces – Mesoporous material – Rotational diffusion – Raman spectroscopy

## 1. INTRODUCTION

Water is undoubtedly the most important substance on earth. It is ubiquitous in nature and a necessary liquid for the emergence of life[1]. Although by far the most classic liquid encountered in everyday life, water presents many unusual physical properties, which are not yet fully understood[2–5]. A large number of studies have highlighted the crucial role of hydrogen-bonding interactions between water molecules in determining its peculiar liquid structure and physicochemical properties [6–9].

In most frequent situations, water is found as spatially confined or in an interfacial state rather than forming a bulk phase. Confining water in synthetic mesoporous solids is a well-suited method to mimic these natural environments due to their high surface-to-volume ratio, well-controlled pore sizes and geometries, and propensities to be functionalized that allows tunable surface chemistry[10]. In particular, mesoporous silica is widely used in industrial applications such as for environmental remediation and waste water treatment[11], drug delivery[12] or nanofluidics[13,14].

Raman and infrared spectroscopies are recognized experimental tools to probe the H-bonds interactions between molecules[15]. Vibrational properties of water confined into mesoporous silica have been extensively investigated[16–22]. The decrease in pore size induces a disruption of the structure of liquid water and unfavors water arrangements with high connectivity[16,19–21,23]. Moreover, the existence of different H-bond interactions was studied from the dependence of the Raman spectra on the hydration level of Vycor porous matrices[22].

This raises the question of the influence of the nature of the pore surface on the interfacial properties of water. This point has been addressed recently by quasi-elastic neutron scattering (QENS)[24] and dielectric relaxation spectroscopy (DRS)[25] studies of the dynamics of liquid water, based on the use of periodic mesoporous organosilicas (PMOs) with carefully designed surface chemistry. The QENS study showed the influence of the nature of the pore surface on the long-time dynamics of interfacial molecules, while the dynamics of water confined in the pore center was barely affected by the interface[24]. Concomitantly, the DRS study has showed that, unlike water in saturated pores, the adsorbed layer formed a reduced relative humidity was most sensitive to water-surface interactions[25]. These findings illustrate the respective importance of water-water and water-surface interactions in determining the dynamics of the interfacial and confined water molecules. They also point out that understanding H-bond interactions is crucial to predict properties such as the molecular dynamics of confined molecules.

Hence, we propose in this article to study the H-bond network of water molecules in the O–H stretching region by Raman vibrational spectroscopy for two different loading situations: (i) pores entirely filled with capillary water and (ii) unsaturated pores with water forming a surface layer leaving the pore center empty. It allowed us to discern populations of water molecules that result from different water/water and water/surface interactions. Varying the filling fraction provided the control of the respective contributions from the interfacial (water/vapor and water/solid) and the pore center regions. Variable temperature experiments were performed to evaluate the strength and the nature of the interactions. In order to vary the surface interaction, we used two different matrices (B-PMO and MCM-41) having comparable pore geometry but different composition. MCM-41 is a prototypical hydrophilic mesostructured silica with cylindrical porous channels. In addition, B-PMO offers a unique opportunity to study the influence of surface chemistry. Its pore wall is formed by a periodic repetition of inorganic (silica) and organic (phenylene) bridging units, with tunable hydrophilicity.[26,27] Unlike post-synthesis surface grafted porous silicas,[28] B-PMO allows a stoichiometric control of the periodically alternating surface chemistry along the pore channel (i.e. one organic bridge per silica inorganic unit) with a repetition distance of 0.75 nm.[29]

## 2. MATERIALS AND METHODS

### 2.1. Materials.

Periodic Mesoporous Organosilicas (PMOs) powders were prepared according to the following procedure. NaOH and the octadecyltrimethylammonium bromide (OTAB) surfactant were dissolved in deionized water. The bis-silylated precursors of the form $(EtO)_3Si-B-Si(OEt)_3$ (B = phenylene unit (-$C_6H_4$-)) were added at room temperature, and the mixtures were stirred for 20 hours. The mixtures were transferred into a Teflon-lined steel autoclave and statically heated to 95 °C or 100 °C for 24 h. The resultant precipitate was collected by filtration and washed with 200 ml deionized water. After drying at 60 °C, the powder was extracted with a mixture of ethanol and hydrochloric acid (EtOH:HCl (37 %), 97:3, v/v) using a Soxhlet extractor. The porosity and the pore structure of the dried materials were characterized by powder X-ray diffraction and nitrogen physisorption.[30]

The mesoporous materials MCM-41 silicas were prepared according to a procedure similar to that described elsewhere[31] and already used in previous works.[32–34] Cetyltrimethylammonium bromide (CTAB) was used as template to obtain hexagonally ordered cylindrical pores, as confirmed by nitrogen adsorption, transmission electron microscopy and neutron diffraction. By studying these matrices, we can study the effect of hydrophobicity/hydrophilicity as illustrated in Table 1.

*Table 1. Structural and chemical parameters of the mesoporous matrices. The hydrophilic and hydrophobic regions are highlighted respectively by red and yellow boxes.*

| Matrix label | Organic bridging unit | Chemical formula | Repetition distance (nm) | Pore surface (m$^2$.g$^{-1}$) | Pore volume (cm$^3$.g$^{-1}$) | Pore diameter (nm) |
|---|---|---|---|---|---|---|
| MCM-41 | … | 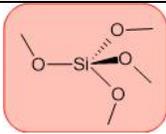 | … | 1077 | 0.89 | 3.65 |
| B-PMO | Benzene | 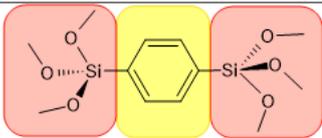 | 0.75 | 815 | 0.68 | 4.1 |

Apart from different surface interactions due to different chemistry, and slightly different pore diameters (3.65 and 4.1 nm respectively) and it is worth noting that MCM-41 and B-PMO present very similar mesoporous structures. They are formed by parallel cylindrical channels arranged on a 2D crystalline honeycomb-like hexagonal array, as illustrated Figure 1. For B-PMO, the surface chemistry of the wall alternates hydrophilic silica inorganic units and hydrophobic aprotic phenylene organic bridges (yellow regions) along the pore axis, with repetition distance 0.75 nm, while MCM-41 comprises only silica units.

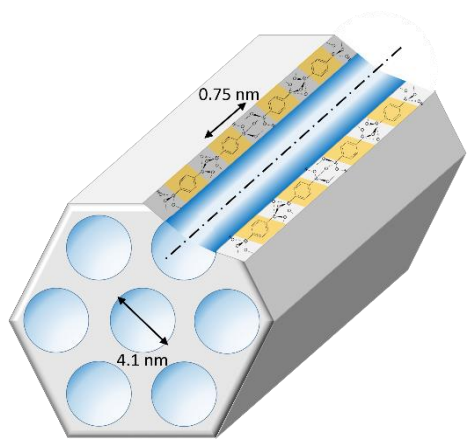

Figure 1. Sketch of the honeycomb-like structure of the mesoporous hosts. For B-PMO, the surface chemistry of the wall alternates hydrophilic silica inorganic units and hydrophobic aprotic phenylene organic bridges (yellow regions), while MCM-41 comprises only silica units.

### 2.2. Methods.

To prepare hydrated matrices, it was found preferable to impose the relative pressure rather than the mass fraction of water that fills the porous materials. It ensures that all the different water-filled materials result from the equilibrium with water vapor at the same chemical potential, which makes their comparison more reliable from the thermodynamic point of view. The filling procedure that was applied in this study can be considered as an experimental realization of the grand canonical thermodynamic ensemble, where the chemical potential of water molecules in the gas phase is imposed by the relative humidity (i.e. RH = $P_{water}/P_{sat}$). At a given temperature, the relation between the amount of water that fills each porous medium and the relative pressure of the saturating atmosphere is determined by the water-vapor adsorption isotherm. They were measured for series of PMOs that were similar to the PMOs used in the present study.[29] They were all characteristic of type V isotherms[35]: at low relative pressure, the isotherm exhibits a low adsorption region, followed by a pore capillary condensation step at an intermediate relative pressure (0.5-0.6, depending on the PMO), and reaches a plateau where the amount of the water increases only slightly with the increase of pressure.

Based on this principle, the matrices powders were prepared in an aluminum cell, and then placed in a desiccator in the presence of a beaker containing a saturated aqueous solution of $MgCl_2$ or NaCl, and equilibrated at around 20 °C for 24 h. The resulting relative humidities (RH) were 33% for the $MgCl_2$ and 75% for the NaCl. The former RH is located below the onset of the capillary condensation, which results in water being adsorbed at the pore surface with an empty pore center. The second RH is above the capillary condensation, leading to the complete filling of the porosity with no excess bulk-like liquid.[29] After the loading procedure, the cell was hermetically sealed.

In principle, water molecules adsorb inside the pores and also outside the grains. However, due to large surface-to-volume ratio of nanoporous regions compared to particles, the amount of water adsorbed outside the pore is negligible. According to SEM images (see Fig. S6 in ref.[30]), the grain size of the mesoporous materials used in the study exceeds a typical size $D_{grain}$ = 200 nm.[30] Assuming spherical shape, the external grain surface is

$$S_{outer} = \frac{6}{D_{grain}}\left(V_{pore} + \frac{1}{\rho_{wall}}\right)$$

where $V_{pore}$ and $\rho_{wall}$ are respectively the pore volume and density of the solid materials forming the matrix walls. Based on previous structural study,[30] the values of $\rho_{wall}$ range from 1.66 g.cm$^{-3}$ to 2.03 g.cm$^{-3}$, and, as indicated in Table 1, the porous volumes $V_{pore}$ range from 0.68 to 0.89 g.cm$^{-3}$ for B-PMO and MCM-41. For both matrices, the outer pore surface is therefore estimated $S_{outer} \approx 40$ m$^2$.g$^{-1}$, which corresponds to less than 5% of the inner pore surface given in Table 1.

Raman spectra were collected in the 2600–4000 cm$^{-1}$ frequency range using a LabRAM-HR800 (Horiba) Raman-spectrometer. A 532 nm laser diode line (39.5 mW) was used as the excitation source. This exciting radiation was focused on the surface of the sample via an Olympus x50 ULWD, 10.6 mm long working-distance objective. Under this condition, the resulting power of excitation laser at the sample was about 11 mW. On cooling, crystallization temperatures were in agreement with previous dielectric spectroscopy and DSC studies, which indicated no significant heating of the sample induced by the excitation laser.[24,25] The scattered light was collected in backscattering configuration and the Rayleigh scattering was removed by means of dielectric edge filters. To remove ripples generated by these filters, we have applied an intensity correction function on the experimental spectra. The cells containing the sample were placed in a THMS600 Linkam temperature device for analyzing the temperature dependence of Raman spectra. Acquisition times were 10 minutes per spectrum for 75% RH-loaded samples and 20 minutes per spectrum for 33% RH-loaded samples. Spectra were recorded for each sample as a function of temperature at a fixed position. To facilitate a direct comparison between spectra, unless specified, each spectrum was normalized to the maximum intensity of its principal component.

## 3. RESULTS AND DISCUSSIONS

### 3.1. Bulk water and saturated hydrophilic MCM-41 pores loaded at 75% RH.

We consider the case of capillary filled MCM-41, which were loaded at a high relative pressure (75% RH). According to the water physisorption isotherms,[29,36] this condition ensures being situated above the capillary condensation filling. Thus, the entire porous volume is saturated with water.

Raman spectra of bulk water and water-filled MCM-41 recorded at 20 °C are presented in Figure 2. Special attention was paid to the O-H stretching region (2800 – 4000 cm[-1]) in order to probe the molecular structure[7,8] and the intermolecular H-bonds[19,37]. The Raman signal of the silica matrix was negligible compared to the signal of the confined water, as already observed for other molecules confined in SBA-15 matrix[38,39]. This is in contrast with ref. [40] that reported prominent contributions from laser radiation-induced defects. Spectra of bulk water and confined water exhibit qualitatively the same shape, composed of a broad triple hump extending over this selected spectral region. This observation is in agreement with previously reported studies on water confined in various silica matrices[16,19,20,40]. The first component (~3200 cm[-1]) is generally associated to the O-H stretching vibration of water molecules involved in a tetrahedral structure, while the second (~ 3400 cm[-1]) corresponds to the distorted H-bond network[41]. The third component (~ 3600 cm[-1]) is related to O-H stretching of water molecules that are not involved in intermolecular HB (named free water).

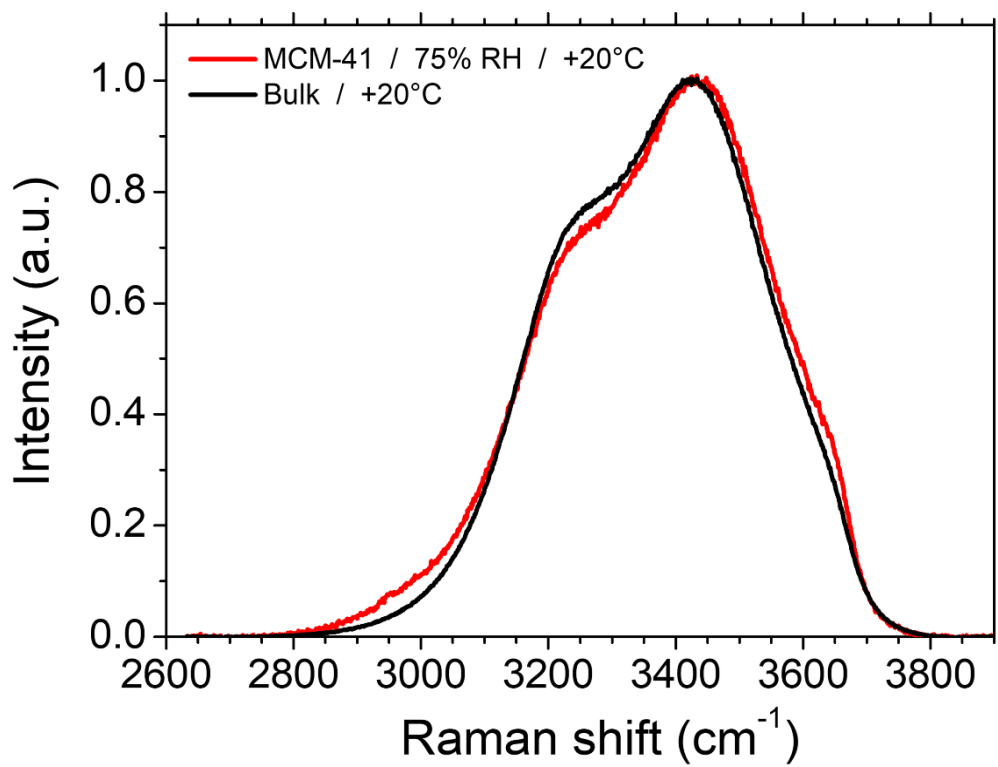

*Figure 2. Raman spectra in the O-H stretching region of bulk water and water-filled MCM-41 recorded at 20 °C.*

The temperature dependence of the Raman spectra was investigated on the temperature range from 20°C to -140°C, as shown in Figure S1. During cooling, crystallization of water in the bulk and confined states was identified by a sudden change of the shape of the spectra, and the emergence of a sharp line around 3100 cm⁻¹, which was otherwise absent in the liquid phase (cf. Figure S1). The occurrence of this additional 'ice-peak' agrees with previous Raman freezing-melting studies[40]. Due to confinement effects, crystallization started below -40°C in water-filled matrices, in agreement with previous dielectric spectroscopy and DSC studies.[24,25] Compared to bulk ice, the spectra of water-filled matrices measured below the freezing point, e.g. at $T$ = - 100°C (cf. right lower panel in Figure S1) present a blue-shifted excess of spectral intensity above 3400 cm⁻¹. A residual "free-water peak" is even noticeable at about 3600 cm⁻¹. This indicates that in confinement, ice is defective and co-exists with amorphous regions, in agreement with previous studies indicating the existence of an unfreezable interfacial layer. [25]

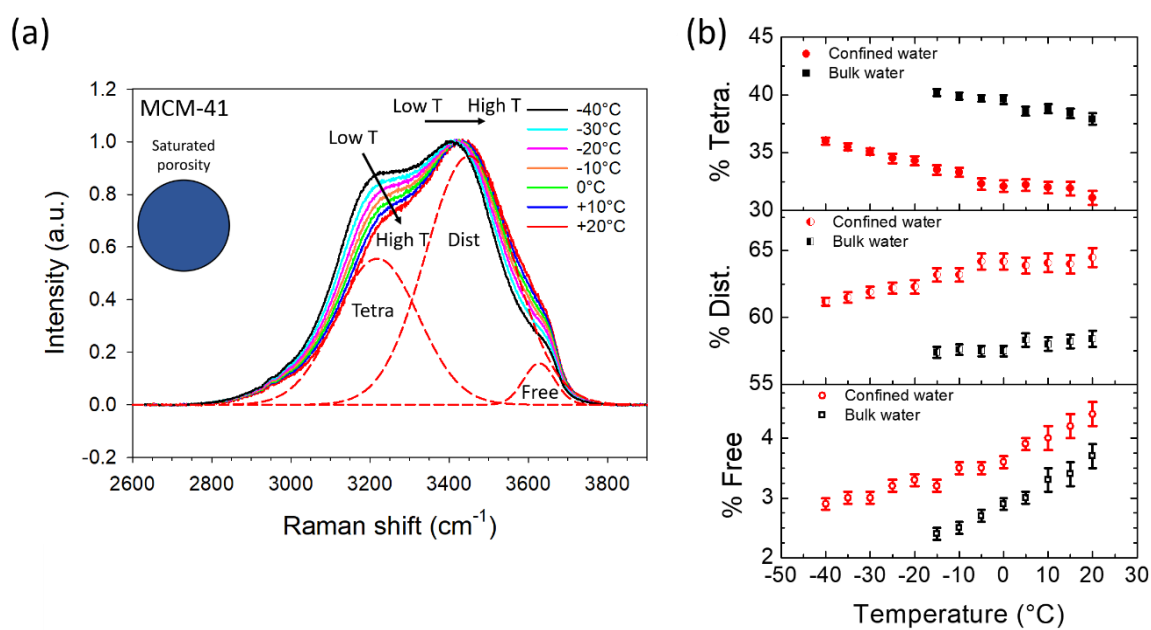

Figure 3. (a) Raman spectra in the O-H stretching region of confined water into MCM-41 at different temperatures. Fitted bands corresponding to the three populations of water molecules at 20°C (dashed lines) (b) Temperature dependence of the three different populations of the water structure for bulk water and water confined in MCM-41. Error bars are standard deviations of the parameters evaluated from the fitting procedure.

The present study focusing on the structure of the liquid state, we restricted the spectral analysis to temperatures above -15°C and -40°C for the bulk and capillary-filled samples, respectively. The corresponding spectra are presented in the Figure 3a for water-filled MCM-41. They revealed systematic variations of the intensity and frequency position of the different bands with the temperature, as indicated by arrows in Figure 3a. To get a quantitative description, a sum of three Gaussian functions (dashed lines in Figure 3a) was fitted to each experimental spectrum (Figure S2), following the scheme applied in previous studies[16,17,19,42–44]. The temperature dependence of the fraction of each population is plotted in Figure 3b. It was defined as the ratio of the integrated intensity of each specific band over the sum of the three integrated intensities. Upon decreasing temperature, the fraction of H-bonded water in tetrahedral environment increases at the expense of the other two populations. This observation reflects the stabilization/disruption of the H-bond network with decreasing/increasing temperature.[45] Compared to bulk water, a qualitatively similar temperature dependence was observed for water-filled MCM-41. As shown in Figure 3b, the study of confined water could be extended to lower temperature without crystallization (down to -40°C) in agreement with a DSC study.[24] However, the fractions of distorted H-bonded and free water are systematically larger in the confined state. This indicates that confinement affects the structure of liquid water and reduces the formation of tetrahedral arrangements[20,23]. Crupi *et al.*[20] have also reported the disruption of the structure of water upon confinement in hydrophilic nanoporous GelSil with a pore size of 7.5 and 2.5 nm, with greater effect for smaller pore size.

Additionally, we observed a slight shift of the three bands towards high frequency in confinement. This observation was made at all the studied temperatures.

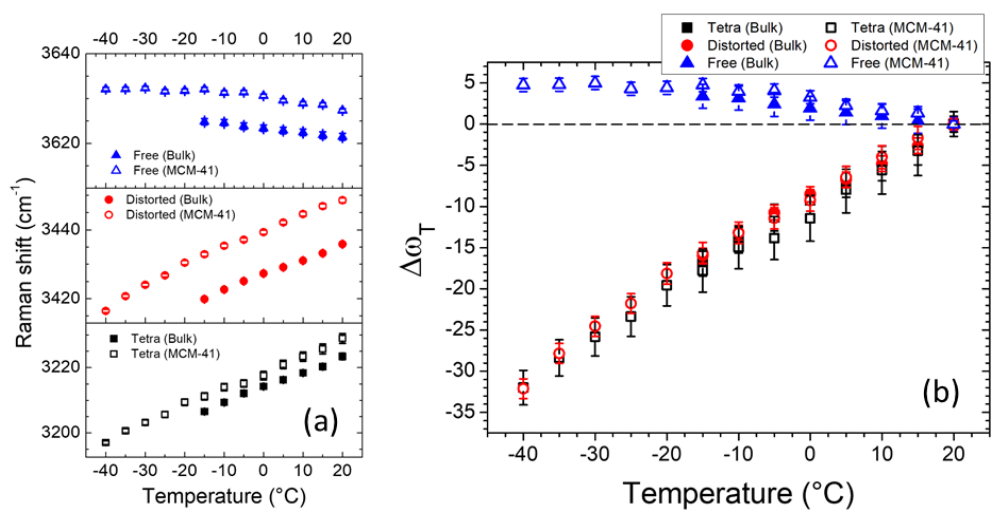

*Figure 4. (a) Temperature dependence of the band positions obtained by the fitting procedure of the intramolecular O-H stretching in bulk water and confined water. (b) Temperature dependence of $\Delta\omega_T = \omega_T(T) - \omega_T(T = 20°C)$. Solid lines are guides to the eyes.*

A similar blue shift was obtained in previous study of water-filled MCM-41 and it indicates a weakening of the interaction strength in confinement.[40] It is clearly visible in Figure 2 for the spectra measured at 20°C. The temperature-dependence of the band positions of the three water populations obtained from the fitting procedure for bulk water and confined water in MCM-41 are illustrated in Figure 4a. Also their difference with respect to the frequency position at $T$ = 20°C $\Delta\omega_T = \omega_T(T) - \omega_T(T = 20°C)$ is presented in Figure 4b. The temperature dependence of these bands gives clear information on the intermolecular interactions of water molecules. Indeed, the positions of the bands assigned to water involved in tetrahedral and distorted H-bond network shift toward high wavenumbers on increasing the temperature, which is the unambiguous signature of vibrational bands involved in H-bonding[45]. Contrariwise, the free O-H band slightly shifts toward the low wavenumbers, indicating that this vibrational band is not involved in H-bond interaction. The frequency shifts $\Delta\omega_T = \omega_T(T) - \omega_T(T = 20°C)$ (Figure 4b) exhibit almost linear temperature variations with slopes that are the same for the two types of H-bonded water populations (i.e. tetrahedral and distorted configurations) and for the two physical states (i.e. bulk and confined water). It suggests that the nature of the H-bond interactions is the same for these different populations, in other words water-water interactions. The value of the slope evaluated by linear regression (0.52 ± 0.01 cm$^{-1}$/°C) is consistent with previous work[8].

### 3.2. Saturated hydrophobic B-PMO pores loaded at 75% RH.

We now extent the discussion to the case of water confined in a B-PMO hydrophobic porous matrix. In the four panels of Figure 5 the Raman spectra of water-saturated B-PMO recorded at different temperatures ranging from -40°C to 20°C are compared with those of water-filled MCM-41. As for MCM-41, this temperature range is located above the freezing point of water in B-PMO, the latter having been identified by the emergence of the 'ice-peak' at 3100 cm$^{-1}$ (cf. Figure S1). Unlike water-filled MCM-41, sharp bands are observed in the spectral region 2800 – 3200 cm$^{-1}$ for water-filled B-PMO. They are characteristic of C-H vibrations[43,45] and assigned to the organic bridging unit of the B-PMO matrix. These additional contributions from the matrix are sharp and relatively intense, which impeded a fit of the entire spectral range. However, their location in the low frequency range of the spectral range allowed to disentangle most of the water contribution, which is located above 3200 cm$^{-1}$. As shown in Figure 5, the signal related to water confined in B-PMO is qualitatively similar to that of water-filled MCM-41. It is attributed to the three populations of water. At 20°C, apart from the additional lines due to organic bridges, the spectra of water-filled B-PMO and water-filled MCM-41 are really superimposed. However, deviations appear on cooling. In B-

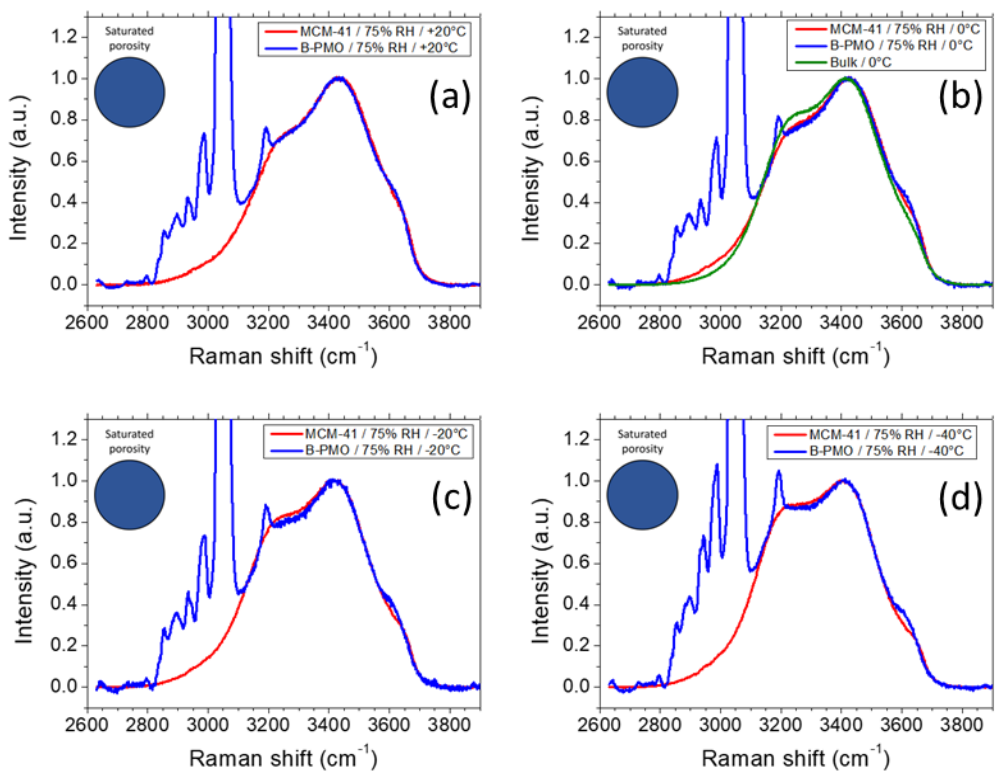

*Figure 5. Raman spectra in the O-H stretching region of confined water in fully filled MCM-41 (red line) and B-PMO (blue line) recorded at (a) 20 °C, (b) 0 °C, (c) -20 °C and (d) -40 °C.*

PMO, it consists in an increase of intensity at about 3600 cm$^{-1}$ and, to a lesser extent, a reduction of intensity around 3200 cm$^{-1}$. The first band corresponds to the population of free O-H and the latter to water molecules in tetrahedral environment. The spectra being normalized to maximum intensity at about 3400 cm$^{-1}$, the relative populations of water in tetrahedral and distorted environment is inaccessible without a complete fit. However, we can formulate the general conclusion that the formation of H-bonds between water molecules is more inhibited in B-PMO than in MCM-41. This observation can be related to the surface chemistry of B-PMO. The existence of one hydrophobic aprotic benzyl organic bridge per hydrophilic silica inorganic unit reduces the overall water-surface interaction. Moreover, the periodically alternating interaction along the pore channel may additional disrupt the formation of an extended H-bond network within water. This interpretation is supported by multidimensional solid-state NMR study[29] of these materials. Strong correlations between water and the surface of the pore were observed in the vicinity of inorganic silica units that present H-bonding silanols groups[29]. On the contrary, a reduction of the interfacial correlations was observed in regions located around hydrophobic organic bridging units. While the decrease in temperature normally increases the fraction of H-bonded molecules at the expense of free O–H, we conclude that this phenomenon is reduced in water-filled B-PMO for the fraction of molecules located in regions near organic units, which are depleted in H-bonding sites.

Crupi *et al.*[20] have observed a weak variation of the relative integrated intensity of the bands for water confined in hydrophilic nanoporous GelSil glasses (pore size 7.5 and 2.5 nm). However, they reported a recovery of bulk-like parameters for water-filled Si–CH$_3$ grafted hydrophobic nanoporous GelSil (pore size 2.5 nm). In the present study, the observations made for water-filled B-PMO definitely follow a different trend. Indeed, at given temperature (see Figure 5 at $T$ = 0 °C), the fraction of free O-H at about 3600 cm$^{-1}$ increases according to the sequence: bulk, MCM-41, and B-DVB. GelSil glasses and PMOs differ in pore geometry (disordered vs mesostructured with crystal-like organization of the organic bridges within the wall of cylindrical channels) as well as in the methods used for chemical functionalization (post-synthetic grafting vs condensation reactions of bridged organosilica precursors), which result in different confining conditions. Moreover, a specificity of B-PMO stands in the periodic modulation of hydrophilic/hydrophobic sites along the channel axis that might additionally disrupt water structure.

### 3.3. Hydrated pore surface of hydrophilic MCM-41 loaded at 33% RH.

The spectral signature of water molecules adsorbed on the surface of MCM-41 was studied for samples that were hydrated at a reduced relative pressure (33% RH). According to the water physisorption isotherms,[29,36] this condition ensures being situated before the capillary condensation filling. Thus, only a layer of adsorbed water can be formed on the pore wall, leaving the pore center empty. Moreover, the contribution from water molecules adsorbed outside the pore onto the surface of the grain is estimated to be less than 5% (cf. Materials part), and can thus be neglected.

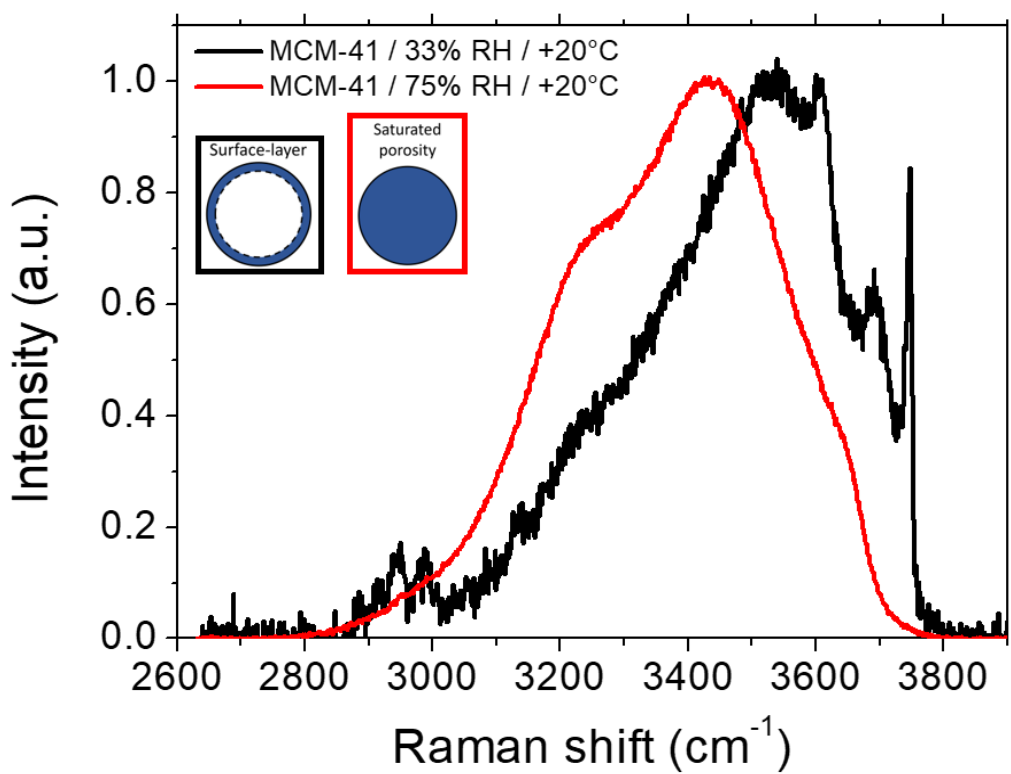

*Figure 6. Raman spectra of confined water in MCM-41 in the two loading situations, loaded at 33% RH below the capillary condensation (black line) and loaded at 75% RH above the capillary condensation (red line).*

Raman spectra of confined water in MCM-41 for the two loading situations (capillary filled and surface-layer) are plotted in Figure 6. We recall that the spectra shown in Figure 6 were scaled to unity at maximum intensity for a better identification of the shifts of the different bands. In fact, the measured spectra significantly differed in their respective total intensity due to the different amount of water, as shown in Figure S3. Moreover, significant differences in the shape of the spectra are observed. Firstly, the entire spectrum is shifted to higher frequency, with a maximum position going from about 3400 cm$^{-1}$ to 3550 cm$^{-1}$ as the RH is reduced from 75% to 33%. This is the typical signature of an overall weakening of the H-bond interactions. Moreover, additional bands can be detected in the spectrum of the surface-layer sample. They suggest that additional contributions from water or silanol hydroxyl groups are involved, opening possibilities to assess the interactions at the water/surface and water/vapor interfaces.

The fitting procedure is presented in Figure 7a for the spectrum recorded at $T =$ 10 °C. The fits obtained at different temperatures are presented in Figure S4. In the spectral region corresponding to OH vibrations (3000-3800 cm$^{-1}$), a minimum of six components was required to achieve an acceptable fit of the O-H stretching spectrum. An alternative procedure with only five bands is also illustrated in Figure S5 and it clearly shows the need for an additional component to describe the bimodal shape of the sharp line around 3750 cm$^{-1}$. It can be also mentioned that two weak bands were observed about 2940-2960 cm$^{-1}$, i.e. located below the spectral range of interest. Prominent peaks located at similar positions were reported in MCM-41 fully saturated with water[40]. They were attributed to the mesoporous matrix, and possibly caused by laser radiation–induced defects. In contrast, in the experimental conditions we used in the present study, the contribution from the matrix was considerably reduced. As illustrated in the Figure 6, these two peaks are weakly detected for partially-filled systems. They are hardly visible for water-filled matrix, having simply been overwhelmed by additional intensity coming from other water molecules, as demonstrated by the unscaled Raman spectra in Figure S3.

In a previous dielectric spectroscopy study, it was shown that the water layer adsorbed on the surface of partially filled MCM-41 and B-PMO remained non-freezable down to the lowest studied temperature (-120°C)[25]. Thanks to the suppression of crystallization, we could perform variable temperature investigations of the water layer on an extended temperature range (from 20°C to -120°C). It provided valuable information on the nature of intermolecular interactions and thus allowed us to assign all bands of the spectrum.

Importantly, even at the lowest temperature studied, we did not find any evidence for the appearance of the band at about 3100 cm$^{-1}$, often referred to as ice-peak. This fingerprint of ice-like order has been observed in totally filled SBA-15 and MCM-41 with pore diameter ranging from 8.9 to 2.0 nm[40]. For weak confinement, it was attributed to crystalline ice. For strong confinement (e.g. 2 nm), despite the suppression of crystallization, the ice-peak was still observed at slightly blue-shifted frequency and attributed to low density amorphous ice[40]. Our results obtained for partially filled MCM-41 point to a very different situation. Indeed, the absence of ice-peak demonstrates that the structure of the non-freezable interfacial layer is definitely different from that of amorphous ice.

The position and assignment of the different bands are summarized in Table 2. The two most intense bands were found at about 3350 and 3550 cm$^{-1}$. These two bands also exhibited the largest temperature dependence in position, as illustrated by the frequency shift $\Delta\omega = \omega(T) - \omega(T = 20°C)$ in Figure 7b. This redshift observed on cooling is the signature of H-bond interactions, and they were attributed to OH vibration of H-bonded water molecules. These bands are however considerably blue-shifted (about +100 cm$^{-1}$) with respect to those due to tetrahedral and distorted H-bond network seen in the bulk and water-saturated MCM-41. Also their temperature dependences (0.38 ± 0.05 and 0.14 ± 0.01 cm$^{-1}$/°C) are smaller than measured for bulk water and water-saturated MCM-41 (0.51 ± 0.01 and 0.52 ± 0.01 cm$^{-1}$/°C). This demonstrates that, at 33 %RH, the adsorbed water molecules are involved in weaker H-bonds. The former band is attributed to water donor molecules involved in water-water H-bonds, and the second band to water donor molecules involved in water-surface H-bonds with oxygens from silanol group or silica as H-bond acceptor.

The position of the third band (3610 cm$^{-1}$) is, within +/- 10 cm$^{-1}$, consistent with the assignment to free water already made for bulk and water-saturated MCM-41. This interpretation is also supported by its temperature variation. Indeed, its position remained constant on the entire temperature range (Figure 7b), which means that it is not related to H-bonds. Moreover, its intensity considerably decreased on cooling, as shown in Figure 8 where the spectra acquired at the two extreme temperatures, -120°C and 20°C, are compared and also for several temperatures in Figure S5. This behavior evidences the decrease of the population of non-bonded water molecules on cooling.

Table 2. Frequency shifts and assignment of the different bands. For temperature dependent bands the value around room temperature and its temperature dependence (in parenthesis) are indicated, while only the mean value on the studied temperature range is given otherwise.

| | Bulk water | Water-filled MCM-41 (75% RH) | | Water-filled MCM-41 (33% RH) |
|---|---|---|---|---|
| | | | **Matrix modes** | |
| | | | unassigned | 2940, 2960 cm$^{-1}$ |
| **Water OH stretching modes** | | | | |
| HO-H→Water (Tetrahedral) | 3220 cm$^{-1}$ (0.51 cm$^{-1}$/°C) | 3230 cm$^{-1}$ (0.51 cm$^{-1}$/°C) | HO-H→Water | 3350 cm$^{-1}$ (0.38 cm$^{-1}$/°C) |
| HO-H→Water (Distorted) | 3440 cm$^{-1}$ (0.52 cm$^{-1}$/°C) | 3450 cm$^{-1}$ (0.52 cm$^{-1}$/°C) | HO-H→Silica | 3550 cm$^{-1}$ (0.14 cm$^{-1}$/°C) |
| HO-H Free | 3620 cm$^{-1}$ | 3630 cm$^{-1}$ | HO-H Free | 3610 cm$^{-1}$ |
| | | | HO-H Dangling | 3700 cm$^{-1}$ |
| | | | **Silanol OH stretching modes** | |
| | | | SiO-H H-bonded | 3740 cm$^{-1}$ |
| | | | SiO-H Free | 3750 cm$^{-1}$ |

Interestingly, the next last three bands at 3700 cm$^{-1}$, 3740 cm$^{-1}$, and 3750 cm$^{-1}$ were absent in capillary filled MCM-41. Therefore, they appear as specific features of the adsorbed water layer. A similar peak near 3700 cm$^{-1}$ has been also found in surface vibrational sum-frequency spectroscopic studies of the water-air interface[46,47]. It was attributed to water dangling hydroxyl bonds pointing towards the vapor phase, which is consistent with the Raman spectrum of the bulk gas phase of water sprays[48]. The same interpretation applies in the present case, if we recognize that, unlike for saturated pores, the water layer adsorbed at 33% RH shares a large internal interface with the empty pore center. It is worth pointing out that the dangling OH cannot be confused with the free OH (cf. band at 3610-3630 cm$^{-1}$ discussed above for bulk water, and for the two water filled MCM-41). Although both situations correspond to non-bonded hydroxyl groups, water-water intermolecular interactions (Van-der-Waals, dipolar) are still present in the case of free OH, leading to a small redshift. In other words, our results demonstrate that in the adsorbed layer of water, non-H-bonded water hydroxyl groups can either point towards the pore center (dangling OH) or within the thin liquid film (free OH).

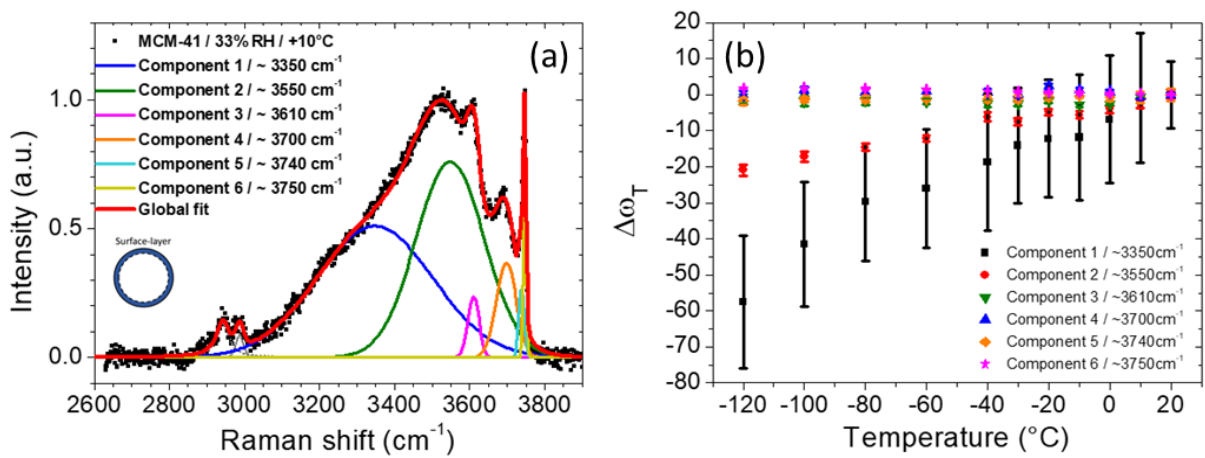

*Figure 7. (a) Fit (thick red line) of a sum of Gaussian peaks (solid thin lines) to the experimental Raman spectra (solid circles) of partially filled MCM-41 at 33% RH corresponding to the formation of a surface water layer. (b) Temperature dependence of the frequency of the different bands.*

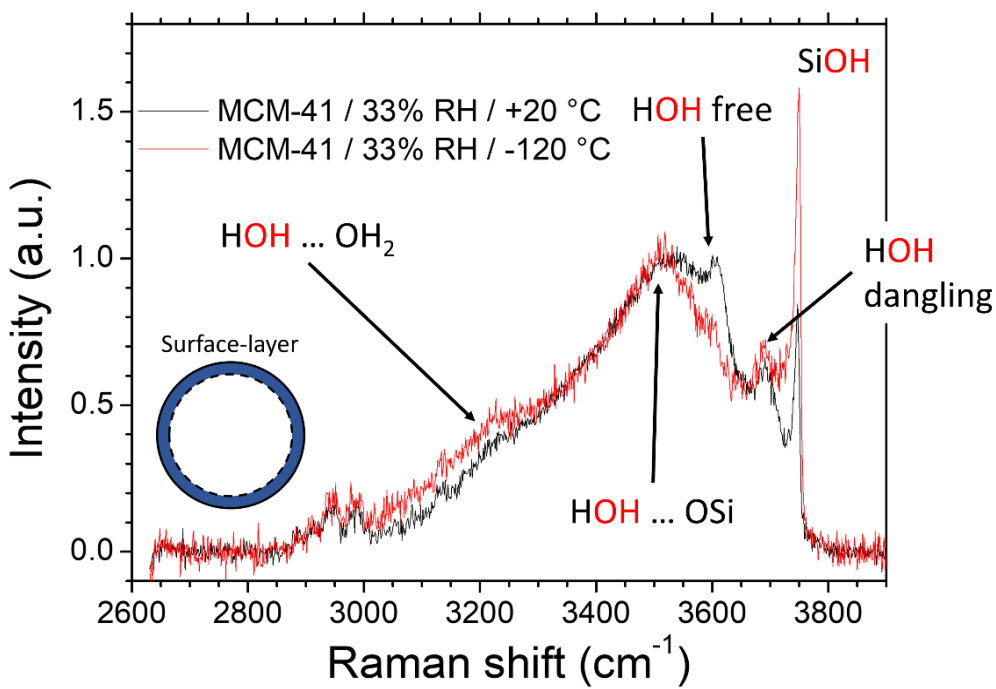

*Figure 8. Comparison of the Raman spectra of partially filled MCM-41 at 33% RH acquired at the two extreme temperatures, -120°C (red line) and 20°C (black line).*

Finally, the last two bands (3740 cm$^{-1}$, 3750 cm$^{-1}$) located at higher frequency than that of dangling OH must be assigned to the vibrational modes of another chemical group. We attribute them to the OH stretching mode of the pore surface silanol groups SiOH. The very narrow band located at 3750 cm$^{-1}$ is characteristic of the free silanol[22]. This line is associated to a slightly red-shifted adjacent line at 3740 cm$^{-1}$. This indicates that different local arrangements co-exist around silanols sites on the pore surface, which may involve the formation of SiOH…OSi weak interactions as well as SiOH…OH$_2$ complexes, with silanol acting as H-bond donor, and adsorbed water as acceptor.

In the literature[22,42], Raman spectra of adsorbed water were reported but restricted to only ambient temperature. Huang *et al.*[22] have investigated different hydration levels for water confined in Vycor glass. The hydratrion level was expressed by the filling fraction $\phi$, defined as the amount of adsorbed water relative to that adsorbed at complete filling. They found similar spectra below filling fraction $\phi$ = 20%, whereas the filling fraction is the present study at 33% RH  was estimated to be $\phi$ = 15% from the water sorption experiment from SI file of ref[29]. A comparative discussion with the results of Anedda *et al.*[42] is hampered by the absence of information about the loading protocol that was applied in that study. We consider that applying variable temperature investigations, as presented in the present study, is a key–parameter to the assignment a vibrational bands and characterization of the nature of intermolecular interactions within the interfacial liquid. It also led us to the important conclusion that, at 33% RH, matrix signal cannot be neglected anymore due to the very small amount of water molecules. The additional bands assigned to OH stretching of silanols provide also usefull information about their contribution to interfacial H-bond interaction with adsorbed water.

Within the surface layer, our results indicate weaker water-water interactions than in bulk or water-filled MCM-41 silica. This is demonstrated by both the smaller magnitude of the red-shift with repect to free water, and also the smaller temperature dependence of the line frequency. According to previous works, water molecules essentially form an adsorbed monolayer film on the surface of MCM-41 at 33% RH [25]. Under these conditions, it is likely that the formation of water H-bonded network is indeed inhibitted due to spatial restrictions and topological constraints induced by the presence of the wall.

Two types of interfacial H-bonds between water and silica are identified, where water is either H-bond donor or acceptor. According to its temperature dependence, the strength of the HO-H→Silica interaction is moderate, but much weaker than the water-water H-bond. Senanayake *et al*[23] have investigated the role of H-bonding role by calculating water Raman spectra by classical MD simulations. They also found that silanols are strong acceptors, but not as strong as water. Unfortunately, in their study, Senanayake *et al*[23] have considered the vibrational bands from water, and not those involving SiOH. Our experiments indicate that silanol may also act as H-bond donor with adsorbed water, though the strength of the interaction is very weak, as quantified by a small (10 cm$^{-1}$) and barely temperature dependent redshift of the stretching mode. This observation somehow contrasts with expectations solely based on the acidic character of SiOH, which might imply that multiple factors actually control the observed temperature dependence of silanols vibrational bands.

### 3.4. Hydrated pore surface of hydrophobic B-PMO loaded at 33% RH.

We now consider the influence of the surface chemistry. The Raman spectra of a water layer adsorbed at 33% RH in B-PMO are compared to those of MCM-41 samples in Figure 9. At first sight, the signal to noise ratio is clearly much smaller for the B-PMO matrix compared to the MCM-41. Such a difference between the two samples was not seen for capillary filled pores. In fact, the two matrices have comparable porous volumes, and so, they contain a similar amount of water at saturation (75% RH). However at 33% RH, a smaller amount of water is adsorbed on the pore surface of B-PMO due to the hydrophobic nature of the organic bridging unit. This conclusion is in accordance with the water physisorption isotherms.[29] Due to the poor quality of the signal to noise ratio in B-PMO, performing a fit of deconvoluted bands to the spectra is challenging, and we rather stick on a qualitative discussion of the most salient features.

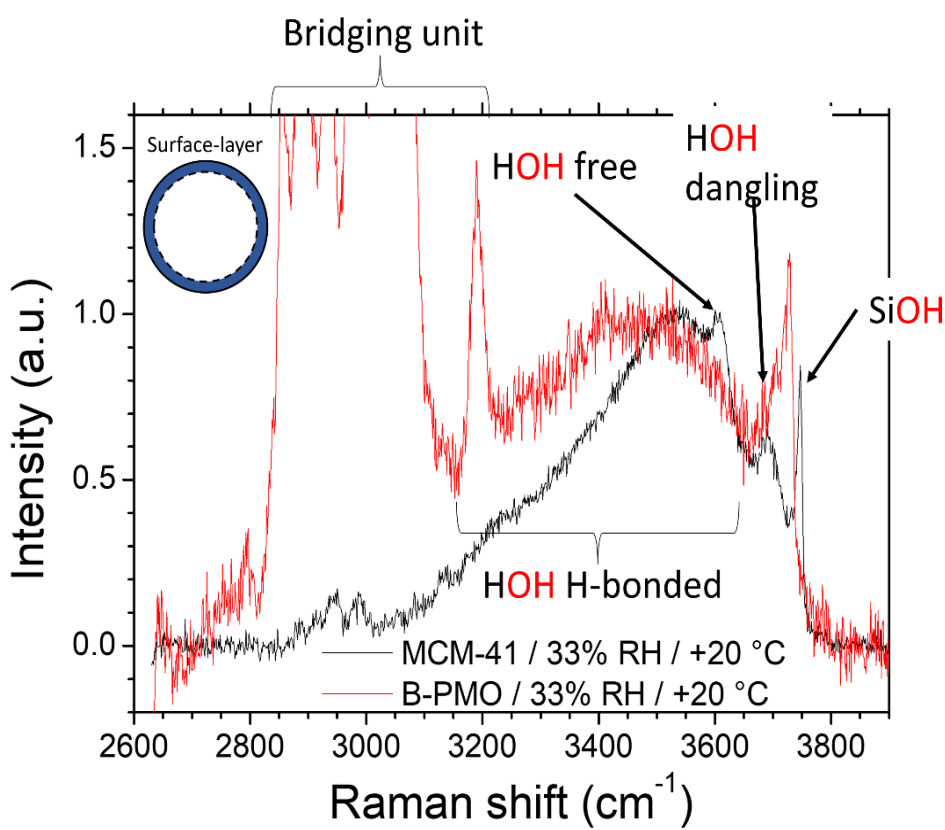

*Figure 9. Comparison between the Raman spectra of a water layer adsorbed at 33% RH in B-PMO (red line) and MCM-41(black line) samples.*

First, it appears that the wide band covering the spectral range from about 3200 $cm^{-1}$ to 3600 $cm^{-1}$ is shifted to lower frequency for B-PMO compared to MCM-41. In MCM-41, this spectral range has been mostly attributed to water-water H-bonds (3350 $cm^{-1}$) and water-silica H-bonds (3550 $cm^{-1}$). In B-PMO, the surface fraction formed by silica units is reduced by the organic linkers that impose a repetition distance of 0.75 nm between inorganic moieties. As a result, one can expect a reduction of the relative intensity of the band related to water-silica H-bonds with respect to water-water H-bonds, which is consistent with the observed apparent redshift. It is worth noting that π-H-bonds between liquid water and benzene have been discussed in the literature, which raises the interesting question about the existence of such interactions in partially filled B-PMO.[49] In that study, the π-H-bond was assigned to a peak at 3610 $cm^{-1}$, which was red-shifted by about 50 $cm^{-1}$ with respect to the dangling OH peak (at 3660 $cm^{-1}$). In fact in Figure 9, the band at about 3600 $cm^{-1}$ is actually seen for the pure mesoporous silica MCM-41, and not for the B-PMO, which rules out its assignment to π-H-bonds in the present case. Such interactions are weaker than those involving water-water and water-silanol partners. As such, they probably barely emerge in the vibrational spectrum, which may explain why we found no evidence of such interactions in our data.

Secondly, the line at 3700 $cm^{-1}$ is much more intense for B-PMO. For MCM-41, we have attributed this band to water molecules located at the interface between the adsorbed film and the vapor phase, pointing a dangling hydroxyl bond towards the pore center. Our observation indicates that the occurrence of this situation is favored in B-PMO. It indicates that the existence of dangling hydroxyl is enhanced by the alternating of spatially distant hydrophilic and hydrophobic sites along the z-axis of the channel. Another phenomenon, also related to the hydrophobic bridging units could additionally contribute the enhancement of the peak at 3700 $cm^{-1}$. Indeed, similar peaks have been reported at a slightly redshifted frequency (in the range 3650-3694 $cm^{-1}$) in vibrational sum frequency spectroscopic studies of hydroxyl groups at a water/hydrophobic-solid interface[46] or at a water/hydrophobic-liquid interfaces (e.g. alkanes, chloromethane)[50]. Similar observation were also made by Raman spectroscopic studies of water in the hydration shell around a hydrophobic solute (neopentane)[51]. In these studies, the marginal redshift with respect to hydroxyl dangling bond at the water/vapor interface has been attributed to weak water-organic interactions that increase with the molecule polarity. X-ray reflectivity studies have also provided evidence for the existence of a hydrophobic gap in the density profile of water at the liquid/solid interface with organically functionalized surface[52]. Similarly for water-filled PMOs, multidimensional solid-state NMR study have concluded on the absence of significant interactions between interfacial water and the hydrophobic organic bridging unit of B-PMO.[29] Therefore, it seems most likely that the dangling hydroxyls identified in the Raman spectra of hydrated B-PMO can be assigned both to the water/vapor interface and to the water/hydrophobic solid interface.

## 4. CONCLUSION

We performed a Raman spectroscopy investigation in the O–H stretching region on confined water within hydrophilic (MCM-41) and hydrophobic (B-PMO) mesoporous matrices. Two different values of filling fractions were achieved by loading from the gas phase and controlling the relative humidity of the cell. The first situation (75% RH, above the capillary condensation) ensures a saturated porosity volume without excess water outside, whereas the second (33% RH, below the capillary condensation) implies that water molecules are adsorbed on the inner surface of the channels leaving the center of the pores empty.

For saturated porosity (75% RH), the spectra related to water confined in MCM-41 and B-PMO are qualitatively similar to those of bulk water. The same classical model could be fitted to all the spectra on an extended temperature range assuming a deconvolution into three populations (*i.e.* H-bonded water molecules in tetrahedral and in distorted environment, and free hydroxyls). From this model, we conclude that confinement limits the H-bond network in liquid water, as shown by a reduction in the fraction water molecules involved tetrahedral network to the benefits of water involved in distorted H-bond environment and free water molecules. This tendency is amplified for hydrophobic B-PMO with respect to hydrophilic MCM-41. However, despite different relative populations, the intermolecular H-bonds probed by Raman vibrational spectroscopy at full loading are dominated by water-water interactions, and so their nature in confinement remains inherently the same as in bulk water.

A very different situation is obtained for partially filled porous matrices (33% RH), where water is present as a layer adsorbed on the pore surface. Two remarkable features of the Raman spectra demonstrate that the structural nature of interfacial water markedly differs from that of water in the bulk state or confined in the core of totally-filled nanopores. Firstly, from the absence of the as-denoted ice-peak, we conclude that the non-freezable water layer does not exhibit the structural fingerprint of low density amorphous ice, even at the lowest temperature studied (-120°C). Secondly, we have identified three additional bands, which we assigned to the specific environments of water hydroxyls groups found at both the water/solid and at the water/vapor interfaces of the adsorbed layer.

For partly hydrated MCM-41 (33% RH), an extended model could be fitted to the spectra. Based on the temperature dependence of the different modes frequencies, we can conclude on a significant reduction of the strength of the water-water H-bond network within the adsorbed film. We show that water also acts as an H-bond donor with oxygens of silica (SiOH or SiOSi), although this interaction is even weaker. Remarkably, our study also points to the existence of two types of non-H-bonded water hydroxyls. The first population corresponds to the so-called free water, as seen in water-saturated pores and more classically seen in bulk water. As in the case of bulk water, the spectral frequency of the free-water hydroxyl mode is slightly red-shifted by non-specific interactions. In other words, these hydroxyl groups are non-H-bonded, but they point to, and interact with, the adsorbed water layer. The second population is ascribed to the dangling hydroxyls of the water molecules at the liquid/vapor interface, which point to the pore center. This is demonstrated by a blue-shifted spectral position that evokes surface vibrational sum-frequency spectroscopic studies of the water-air interface. Finally, silanols also marginally act as H-bond donor.

In partly hydrated B-PMO (33% RH), the relative fraction of water dangling hydroxyls increases with respect to MCM-41, which demonstrates the importance of the surface chemistry and hydrophilicity for the water adsorbed layer. It indicates that the repetition of hydrophilic silica inorganic unit and hydrophobic aprotic phenylene organic bridge reduces the overall water-surface interaction and disrupt the H-bond network within the surface layer. Moreover, we found evidence that for B-PMO, the population of water dangling hydroxyls could relate to the water/vapor interface but also to the water/organic bridge interface as for water/hydrophobic-liquid interfaces[50]. This important conclusion, that finds additional support from multidimensional solid-state NMR study[29], would also imply an increase in dangling hydroxyl groups among the interfacial water molecules on approaching the organic bridging units.

As a whole, this study highlights the differences between the H-bonded structures formed in an adsorbed water layer and at a liquid/solid interface of capillary filled pores, and their dependence on the surface chemistry. It also helps making the relation between the H-bond structures found in these different situations, and the resulting liquid water dynamics as recently reported by QENS and BDS[24,25].

## ASSOCIATED CONTENT

### Supporting Information

Temperature dependence of the Raman spectra of liquid water and ice in the bulk state and in capillary filled MCM-41 and B-PMO at 75% RH (Figure S1); fit of a sum of Gaussian peaks to the Raman spectra of bulk liquid and capillary filled MCM-41 at 75% RH (Figure S2); comparison of the Raman intensity of water–filled samples for two loading conditions (Figure S3); fit of a sum of Gaussian peaks to the Raman spectra of partially filled MCM-41 at 33% RH corresponding to the formation of a surface water layer (Figure S4); comparison of fitting procedures with five and six bands (Figure S5); temperature dependence of the water-related bands in the surface-layer situation in MCM-41 (Figure S6); temperature cycling effect on water-related bands in the surface-layer situation in MCM-41 (Figure S7)

## ACKNOWLEDGEMENTS

This work was conducted in the frame of the DFG-ANR collaborative project (Project NanoLiquids No. ANR-18-CE92-0011-01, DFG Grant No. Fr 1372/25-1- Project number 407319385, and DFG Grant No. Hu850/11-1- Project number 407319385), which is acknowledged. We thank Dr. Malina Bilo for providing the B-PMO sample. We also acknowledge the scientific exchange and support of the Center for Molecular Water Science (CMWS). Raman spectroscopy experiments were realized on the micro-Raman spectrometer of the SIR Platform of the ScanMAT UMS of the University of Rennes 1" (https://scanmat.univ-rennes1.fr).

TOC Graphic

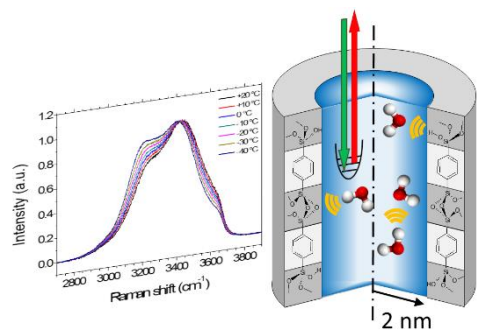

For Table of Contents Only

# Supporting Information for

# Structure of water at hydrophilic and hydrophobic interfaces: Raman spectroscopy of water confined in periodic mesoporous (organo)silicas

Benjamin Malfait[1], Alain Moréac[1], Aïcha Jani[1], Ronan Lefort[1], Patrick Huber[2,3,4,a], Michael Fröba[5,b], and Denis Morineau[1,c]

[1]Institute of Physics of Rennes, CNRS-University of Rennes 1, UMR 6251, F-35042 Rennes, France

[2]Institute for Materials and X-Ray Physics, Hamburg University of Technology, 21073 Hamburg, Germany

[3]Centre for X-Ray and Nano Science CXNS, Deutsches Elektronen-Synchrotron DESY, 22603 Hamburg, Germany

[4]Centre for Hybrid Nanostructures CHyN, Hamburg University, 22607 Hamburg, Germany

[5]Institute of Inorganic and Applied Chemistry, University of Hamburg, 20146 Hamburg, Germany

[a]patrick.huber@tuhh.de

[b]froeba@chemie.uni-hamburg.de

[c]denis.morineau@univ-rennes1.fr

## Table of content

# 1.  Temperature dependence of the Raman spectra of liquid water and ice in the bulk state and in capillary filled MCM-41 and B-PMO at 75% RH

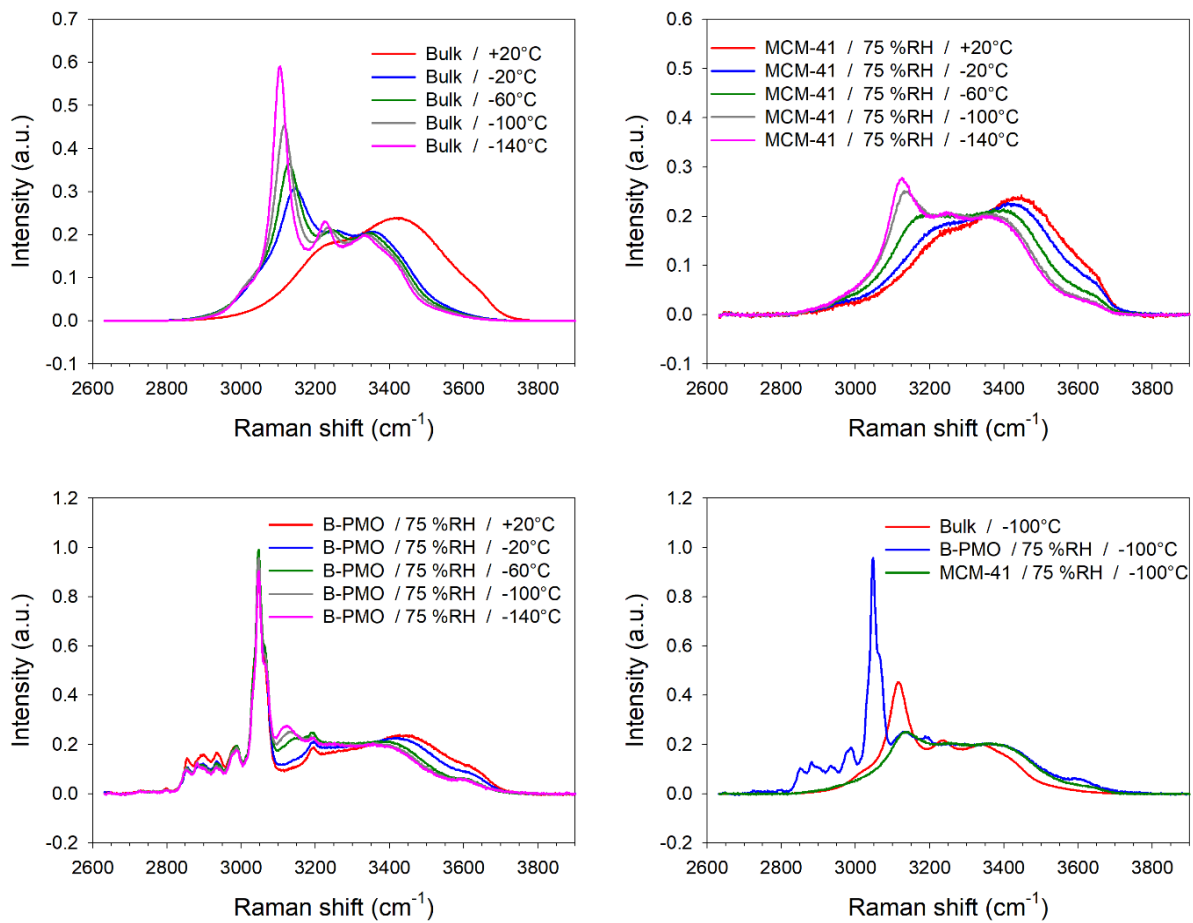

*Figure S1. Temperature dependence of the Raman spectra of bulk water (left upper panel), capillary filled MCM-41 (right upper panel) and B-PMO (left lower panel) at 75% RH that correspond to the saturation of the pores with water. Comparison of the spectra of the three samples at temperature T = -100°C below the freezing point (right lower panel).*

Figure S1 highlights crystallization of water on cooling. In the three conditions studied (bulk, water-filled MCM-41 and water-filled B-PMO) ice formation is identified by a sudden change of the shape of the spectra, and the emergence of a sharp line at 3100 cm$^{-1}$, which is otherwise absent in the liquid phase. Compared to bulk ice, the spectra of water-filled matrices measured below the freezing point, e.g. at $T$ = - 100°C (cf. right lower panel in Figure S1) present a blue-shifted excess of spectral intensity above 3400 cm$^{-1}$. In B-PMO, a residual "free-water peak" is even noticeable at about 3600 cm$^{-1}$. This indicates that in confinement, ice is defective and co-exists with amorphous regions, in agreement with previous studies indicating the existence of an unfreezable interfacial layer.

# 2. Fit of a sum of Gaussian peaks to the Raman spectra of bulk liquid and capillary filled MCM-41 at 75% RH

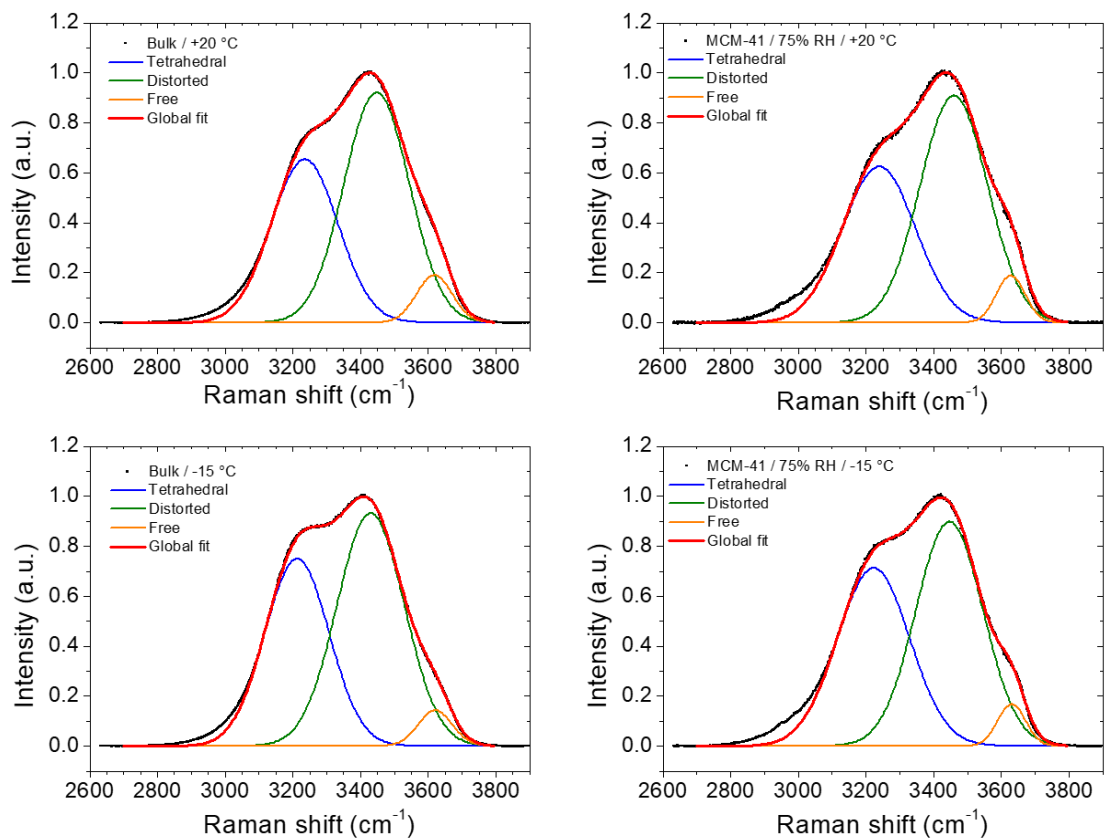

*Figure S2. Fit (thick red line) of a sum of Gaussian peaks (solid thin lines) to the experimental Raman spectra (solid black circles) of bulk water (left panels) and capillary filled MCM-41 (right panels) at 75% RH corresponding to the saturation of the pores with liquid water. From top to bottom, two different temperatures are 20°C and -15°C*

## 3. Comparison of the Raman intensity of water–filled samples for two loading conditions

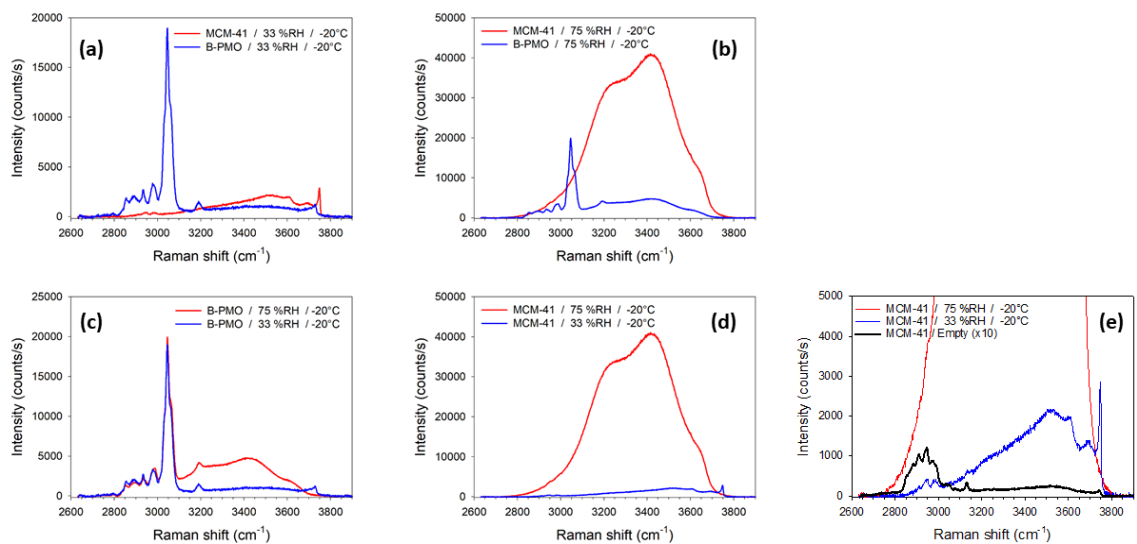

*Figure S3. Comparison of the Raman intensity expressed in count/s of water–filled samples for two loading conditions (33%RH and 75% RH), in MCM-41 and B-PMO at temperature T = -20°C.(a) two matrices at 33%RH, (b) two matrices at 75% RH, (c) B-PMO at two RH, (d) MCM-41 at RH, (e) with rescaled dry MCM-41.*

4. Fit of a sum of Gaussian peaks to the Raman spectra of partially filled MCM-41 at 33% RH corresponding to the formation of a surface water layer

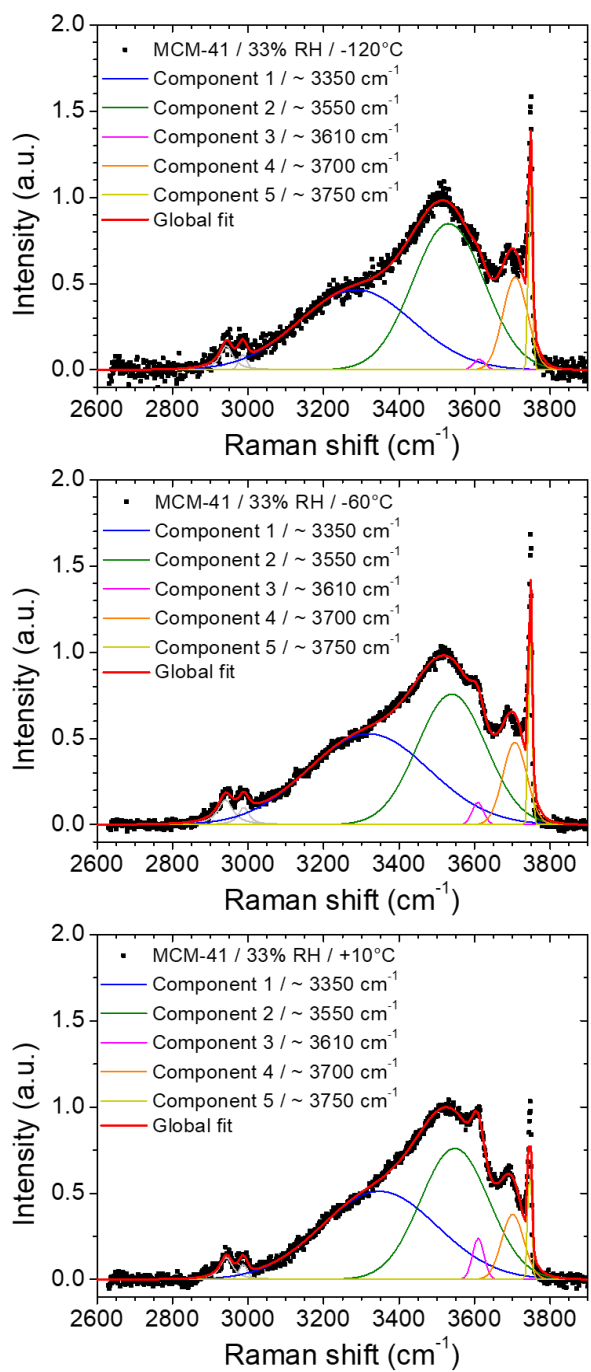

*Figure S4. Fit (thick red line) of a sum of Gaussian peaks (solid thin lines) to the experimental Raman spectra (solid circles) of partially filled MCM-41 at 33% RH corresponding to the formation of a surface water layer. From top to bottom, three different temperatures are -120°C, -60°C, and 10°C.*

# 5. Comparison of fitting procedures with five and six bands

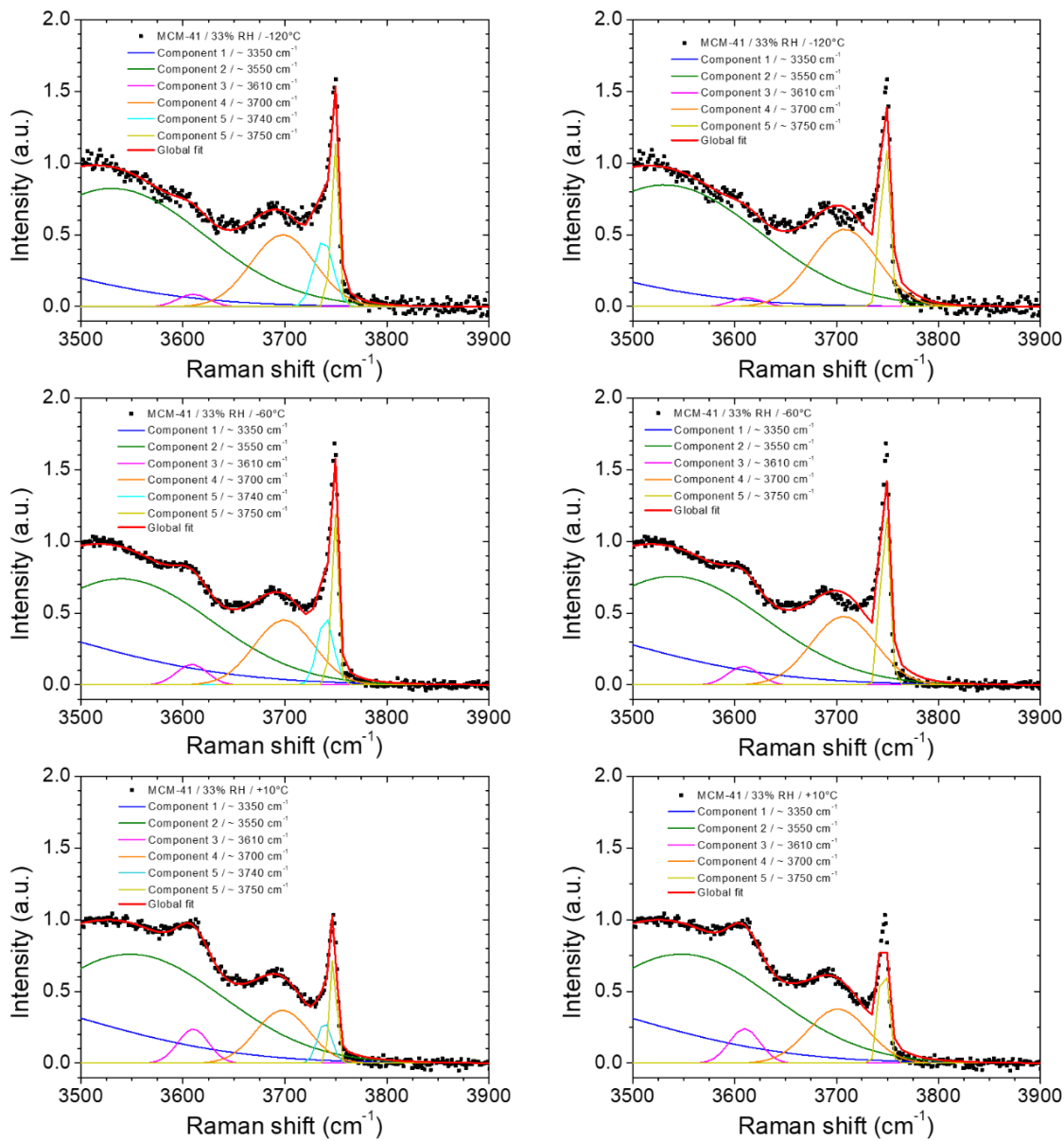

*Figure S5. Comparison of fitting procedures with six (left panels) and five (right panels) bands at different temperatures highlighting the component at 3740 cm⁻¹.*

# 6. Temperature dependence of the water-related bands in the surface-layer situation in MCM-41

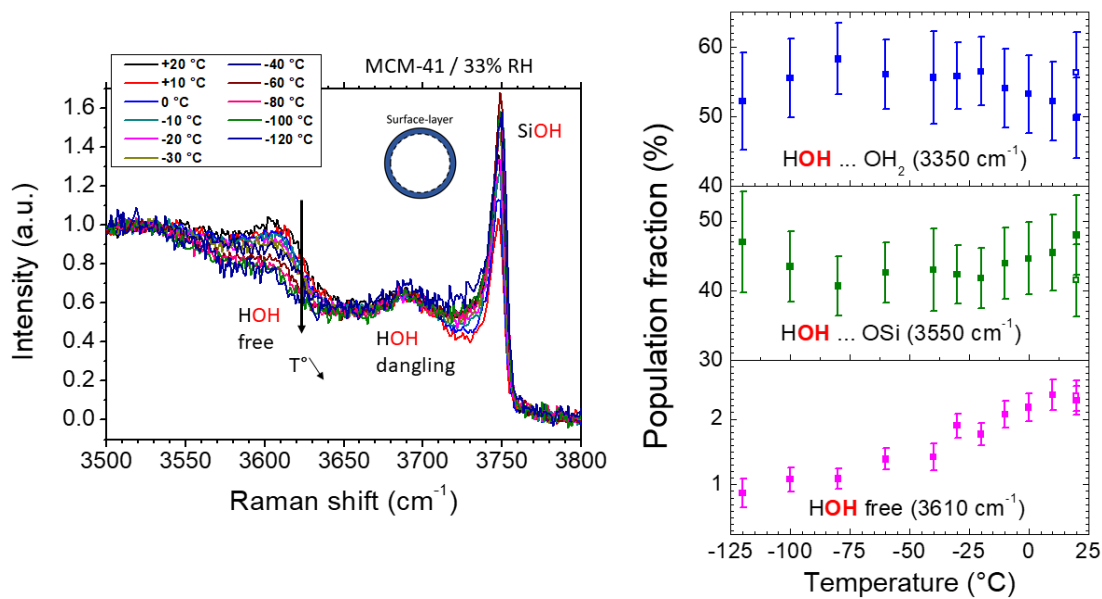

*Figure S6. Left, temperature dependence of (a) the Raman spectra and (b) the three populations of partially filled MCM-41 at 33% RH corresponding to the formation of a surface water layer.*

Figure S6 highlights that, upon decreasing the temperature, the population fraction (left side) of the assigned band O–H free stretching for water molecules in the surface-layer situation decreases. This observation is consistent to a non-H-bonded vibration; the decreasing of the temperature favors the formation of H-bonded population to the detriment of the free O–H.

# 7. Temperature cycling effect on water-related bands in the surface-layer situation in MCM-41

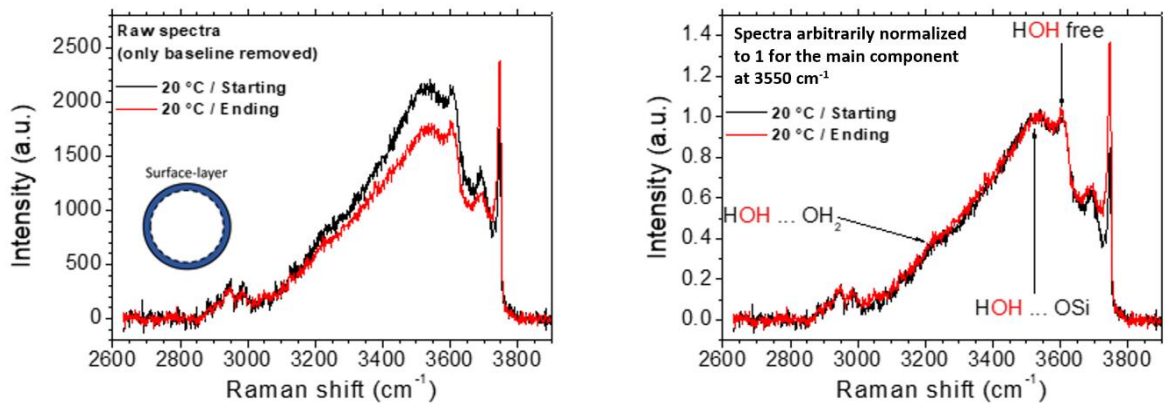

*Figure S7. Spectra of water filled MCM-41 at 33% RH at 20°C measeured at the beginning and at the end of the entire experiement. Left panel: raw data. Right panel: data normalized to maximum water intensity.*

Spectra in Figure S7 indicate that a small fraction of water molecules is lost during the experiment. This occurs probably during the first temperature steps, which are above 0 °C. However, the overall spectrum of water is not modified, as seen Figure S7b after scaling to the maximum intensity of the water modes, which validates the analysis of the temperature dependence of the spectra performed in the study on the basis on the frequency position and on the relative intensity of the different bands.